\documentclass[a4paper,preprint]{elsart}
\usepackage{graphicx}
\usepackage{dcolumn}
\usepackage{amsmath}
\begin{document}

\journal {Chinese Physics} %
\volume{14} %
\pubyear{2005} %
\issue{12} %
\firstpage{2560} %
\lastpage{2564}%
\begin{frontmatter}
\title{A numerical simulation of the backward Raman amplifying in
plasma }
\author{
Wang Hong-Yu$^{1,2}$, Huang Zu-Qia$^1$}
\address {Institute of Low Energy Nuclear Physics, Beijing Normal University,Beijing,100875, China
}
\address {Department of Physics, Anshan Normal
University,Anshan,114005, China }
\date{10/17/2005}
\begin{abstract}

This paper describe a numerical simulation method for the
interaction between laser pulses and low density plasmas based on
hydrodynamic approximation. We investigate Backward Raman Amplifying
(BRA) experiments and their variants. The numerical results are in
good agreement with experiments.

\


\end{abstract}

%\pacs{52.40.Nk,42.65.Yj,52.35.Mw}
\end{frontmatter}
\maketitle

\section{introduction}
Recently, the concept of laser amplifying by backward raman
scattering(Backward Raman Amplifiers,BRA) in plasma was presented
{\cite{Shvets},\cite{Malkin},\cite{Clark},\cite{Fisch}}. In which a
short seed pulse is amplified by a counter-propagating long pumping
pulse when the resonance relation
$\omega_{seed}+\omega_{plasma}=\omega_{pump}$ is satisfied. In the
nonlinear regime, the seed pulse is strongly amplified and
compressed temporally. Theoretically, unfocused intensities of
$10^{17} W/cm^2$ in ~50 fs pulses are accessible by this technique.

Up to now two schemes of BRA have been presented: the amplifier in
non-linear regime and transient regime. The basic concepts of the
first scheme are based on the $\pi$-pulse solutions of laser-plasma
coupled equations{\cite{Malkin}}. When the amplitudes of the plasma
wave and the seed pulse are both quite large, the pumping pulse's
energy is absorbed by the front of the seed pulse. So the front edge
of the seed pulse is amplified while the back edge will be reduced
for the energy feedback into the exhausted pumping pulse. In this
way, the seed pulse is amplified and compressed. Finally, the seed
pulse evolves to a narrow wave train which is called a $\pi$ pulse.

The scheme of transient BRA is based on the matching of the plasma's
density gradient and the pumping pulse's frequency chirp
{\cite{Fisch}}{\cite{Malkin2}}. In the amplifying process, the
strict matching of the two pulses and the plasma frequency is always
needed. If we let the pumping pulse chirp and the plasma density
have a gradient, and set the plasma frequency gradient just be a
half of pumping chirp gradient, the seed pulse will be amplified
only at a fixed position of itself. By passing a plasma of
sufficient length,the amplifying take place and a ultra-fast laser
pulse formed.

Each scheme is very difficult to be realized. For the limit of
plasma technology, the density of gas-jet plasma is limited for
about $10^{19}cm^{-3}$ now. To get distinct amplification, the
seed pulse must travel several millimeters in plasma. The
amplification require rigorous match of the frequencies. Keeping
the uniformity of the plasma is a great challenge in experiments.
In the transient scheme, we need even longer plasma length to
compensate the lower efficiency.

However, some good results of amplifying experiments have been
posted recently{\cite{Yuan}\cite{Cheng}}. Especially, the
experiments informed that the non-linear BRA scheme was possible. In
the experiments by W.Cheng et al, the seed pulses were amplified to
one order higher than pumping pulse and the amplifying ratio was
several thousands.

Consequently, the numerical simulation of BRA becomes a valuable
problem in computational plasma physics. However, compared with the
common plasma simulations there are some special difficulties
because of the long-distance travelling
wave{\cite{Zhang}}{\cite{Xu}}{\cite{clark2}}. First,the computing
time and memory needed is very large. Second, the investigation
requires low-error simulation technology for the low increase rates
of Raman effects . For example, we need a great number of particles
to get rid of the numerical wave-breaking and numerical heating in
Particle-In-Cell method{\cite{Birdsall}}. Besides, we need a
high-order difference scheme of the electromagnetic field equation
to overcome the difficulties of numerical dispersion while applying
it to the long-distance traveling wave. So the simulations of BRA
are mainly focused on two regime now: 1)averaging the effects of
electromagnetic fields to reduce the space-time steps,such as the
APIC method{\cite{Hur}}. 2)using the amplitudes coupled equations
{\cite{Fraiman},\cite{Tsidulko},\cite{Berger}} or linearized plasma
wave equations to get approximately analysis.

An attracting idea is applying a moving window which following the
seed pulse to the computation.This technology is convient to
investigate the interactions between laser and low-density plasma.
In the corona region, the plasma density is about $\frac 1 {10}$ to
$\frac 1 {100}$. The group velocity of laser pulse is very close to
light speed. So the calculated window is shorter than $\frac 1{10}$
of the total interaction region. There is a simulation method which
applies a moving window on the standard PIC simulation (Moving
window PIC Simulation,MWPIC){\cite{Nieter}}. However MWPIC needs the
space-time step match($dx=cdt$) to avoid the interpolation error.
The step setting is unstable at 2-dimensions and 3-dimensions
explicit differential schemes. That means we need some implicit
differential schemes. The latter technology are still developing
now. In addition, MWPIC need a very large number of particles to get
acceptable precision too.

For the above reasons, we developed a numerical simulation technique
based on hydrodynamic approximation. In this technique, we only use
the cold-plasma approximation which is often used in theoretical
researches. The numerical trials indicate that the technique is
suitable for the traveling wave problems in plasmas and can give
results in concordance with experiments. Besides, the technique has
a clear physical picture and the results can be easily analyzed.

We will discuss the BRA equations and the simulation technology in
section 2 and section 3, then analyse the numerical results in
section 4.

\section{the basic equation of BRA}
In the typical instance of BRA, we can introduce the cold-plasma
approximation and the transverse canonical momentum
conservation{\cite{Sprangle}}: $\vec{v}=\vec{u}+\frac
{e\vec{A}}{\gamma mc}=\vec{u}+c \vec{a}$, where $\vec{u}$ is the
longitude velocity and $\vec{A}$ is the vector potential at Coulomb
gauge while $\vec{a}$ is the normalized vector potential:
$\vec{a}=\frac {e \vec{A}}{m c^2}$. Because the electrons'
velocities are far less than the light speed,we can ignore the
relativistic factor $\gamma$.

The plasma hydrodynamic equation can be written as:(We ignore the
thermal pressure here. We  can recover it to the rhs. of the
second equation whenever needed)
\begin{eqnarray}
&\frac{\partial n}{\partial t}+\nabla(n\vec{v})  &= 0 \\
&\frac{\partial \vec{v}}{\partial t}+(\vec{v} \cdot
\nabla)\vec{v}&=-\frac {e\vec{E}}{m}-\frac {e(\vec{v}\times B)}{mc}
\end{eqnarray}

Supposing the plasma flow are non-vertex before the laser action
and applying the cold-plasma approximation,we get the plasma
hydrodynamic equation:
\begin{eqnarray}
&\frac {\partial n}{\partial (ct)}+(\vec{\beta} \cdot \nabla)n &=-n(\nabla \cdot \vec{\beta})-(\vec{a} \cdot \nabla)n \\
&\frac {\partial \gamma \beta}{\partial (ct)}+(\vec{\beta} \cdot
\nabla)\vec{\beta}&=\nabla \phi-\frac 12\nabla\vec{a}^2 -
\nabla(\vec{\beta} \cdot \vec{a})
\end{eqnarray}

The electromagnetic field satisfies the wave equation and Poisson
equation:
\begin{eqnarray}
(\frac 1 {c^2}\frac {\partial^2}{\partial
t^2}-\nabla^2)\vec{a}&=-\frac {4\pi ne^2}{mc^2}\vec{a}
\end{eqnarray}
\begin{eqnarray}
\label{poisson_equ}
\nabla^2{\phi}=\frac {4\pi e^2}{mc^2}(n-n_i)
\end{eqnarray}
Applying the coordinate transformation
\begin{eqnarray}
\xi=x-ct \\
\eta=t
\end{eqnarray}
we get the wave equation in the new variables:
\begin{eqnarray}
(\frac {1}{c^2}\frac{\partial ^2}{\partial \eta^2}-\frac {2}{c}\frac
{\partial^2}{\partial \xi \partial \eta} -\frac {\partial
^2}{\partial y^2}-\frac {\partial ^2}{\partial z^2})\vec{a}=-\frac
{4 \pi n e^2}{mc^2}\vec{a}
\end{eqnarray}
and the Poisson equation has no change.

To solve the wave equation numerically,we introduce the medium
variable $\vec{\psi}$:
$$(\frac 1 c \frac {\partial }{\partial \eta}-2\frac {\partial}{\partial \xi})\vec{a}=\vec{\psi}$$
Finally,we get a pair of first order differential equations:
\begin{eqnarray}
\label{vector_p_equ}
\nonumber \frac {\partial \vec{\psi}}{\partial \eta}=-\frac {4\pi n e^2}{mc^2}\vec{a} +\nabla_{\perp}^2\vec{a}\\
\label{stars}\frac 1 c \frac {\partial \vec{a}}{\partial
\eta}=\vec{\psi}+2\frac {\partial{a}}{\partial \xi}
\end{eqnarray}

With the same transformation,the hydrodynamic equations became
\begin{eqnarray}
\label{fluid_equ} \nonumber\frac {\partial n}{\partial
t}+(\beta_x-1)\frac {\partial n}{\partial x}+\beta_y\frac{\partial
n}{\partial y}=-n[\frac
{\partial (\beta_x -1)}{\partial x}+\frac {\partial \beta_y}{\partial y}]\\
\nonumber\frac{\partial \beta_x}{\partial t}+[(\beta_x-1)\frac
\partial {\partial x}+u_y \frac \partial {\partial y}](\beta_x -1)=
\frac {\partial \phi}{\partial x}-\frac 12 \frac {\partial {\vec{a}^2}}{\partial x}\\
\frac {\partial \beta_y}{\partial t}-[\frac \partial {\partial
x}+\beta_y\frac {\partial}{\partial y}]\beta_y= \frac {\partial
\phi}{\partial y}-\frac 12 \frac {\partial \vec{a}^2}{\partial y}
\end{eqnarray}

The equations
(\ref{poisson_equ}),(\ref{vector_p_equ}),(\ref{fluid_equ}) are the
basic equations to describe the laser-plasma interactions in BRA.

\section{numerical simulation method}
The electromagnetic equations
(\ref{poisson_equ}),(\ref{vector_p_equ}) describe the
electromagnetic wave transmitting in plasma. In BRA scheme, the
transverse dynamic effects are mainly slow processes such as
self-focusing. We can use the periodic boundary conditions in the
transverse directions, apply FFT calculation to eliminate the
transverse derivatives and use direct elimination technology in the
longitudinal direction. So we will focus on  the 1-dimension scheme
at following discussion. However, the generalization to
multi-dimensions is straightforward.

The elimination of Poisson equation is quite simple. In our moving
window, the right boundary of calculating region are always at the
front of seed pulse. We ignore the interactions between the pumping
pulse and the origin plasmas for the uniform of the pumping pulse,
so all laser-plasma interaction can not affect the region in the
right side of the boundary. Then we can assume that there are no
electrostatic fields in the right region of seed pulse: $\vec{E}=0$
then $\phi_i=\phi_{i+1}=...=const$. The differential scheme is
\begin{equation}
\phi_{i-1}=\frac {4\pi e^2}{mc^2}{\delta
x}^2(n_i-N_i)-\phi_{i+1}+2\phi_i
\end{equation}

To get the numerical solution of the wave equation,we use a
generation of R.Liu's leapfrog-upwind scheme{\cite{Liu}}:
\begin{eqnarray}
\nonumber &&\frac {\frac 12 [(a_{i-1}^{n+1}-a_{i-1}^n)+(a_{i}^n-a_{i}^{n-1})]}{\delta t} \\
\nonumber &&-2 c \frac {\frac {1-m}2 (a^{n}_{i+1}-a_i^{n})+m(a_i^n-a_{i-1}^n)+\frac {1-m}2 (a_{i-1}^n-a_{i-2}^n)}{\delta x}\\
\nonumber &=&\frac 1 2c (\psi_i^{n}+\psi_{i-1}^{n})
\end{eqnarray}

Where
\begin{eqnarray}
&\kappa=&2c\frac {\delta t}{\delta x} \\
&m=&\frac 1 6(5+3\kappa-2\kappa^2)\\
\nonumber &a_{i}^{n+1}=&a_{i+1}^{n-1}-(1-2m\kappa)(a_{i+1}^n-a_{i}^n)\\
\nonumber &&+(1-m)\kappa(a_{i+2}^n-a_{i+1}^n)+(1-m)\kappa(a_i^n-a_{i-1}^n)\\
&&+c \delta t (\psi_{i+1}^n+\psi_{i}^n)
\end{eqnarray}

The hydrodynamic equation (\ref{fluid_equ}) can be solved with
leap-frog method too. However,we must apply the FCT technology
{\cite{Zalesak}} to avoid the numerical oscillation while introduce
little numerical damping. The Flux Corrects are used in all
leap-frog steps by rewritten to convection equations:
\begin{eqnarray}
\frac {\partial U}{\partial t}+\frac {\partial F}{\partial x}=S
\end{eqnarray}
F means the convect terms and S means source terms,then the
leap-frog scheme is:
\begin{eqnarray}
 U_{i}^{n+1}=U_{i}^{n-1}- 2 \frac
{\Delta t}{\Delta x}(F_{i+1/2}^n-F_{i-1/2}^n)+2 \Delta t S_{i}^{n}
\end{eqnarray}

To use FCT technology, first we calculate the low-order flux of
plasma density(there is no source term in the continuum equation):
\begin{eqnarray}
&F_{i+1/2}^{low}&=\overline{v_i}*n_i^{DC}*\delta t+\eta (n_{i+1}^{k-1}-n_i^{k-1})\\
&\overline{v_i}&=\frac 12 (v^k_i+v^k_{i+1})\\
&if(\overline{v}>=0)&w_i^{DC}=w^{k-1}_{i}\\
&else &w_i^{DC}=w^{k-1}_{i+1}\\
&w&=n,v
\end{eqnarray}

$\eta$ is a coefficient about $\frac 1 {24}$ to $\frac 1 {8}$. The
smaller $\eta$ means less error but the more numerical oscillation.
Then the high-order flux is:
\begin{eqnarray}
F_{i+\frac12}^{high}=\frac 7{12}(f_{i+1}+f_i)-\frac
1{12}(f_{i+2}+f_{i-1})
\end{eqnarray}
We don't introduce the correct of the velocity's flux because it
will cause much damping in the simulation. So the velocity flux can
use the same scheme as the high-order density flux.

However, the difference scheme of plasma density does not preserve
positivity. To get rid of the difficulties of negative electron
density, we apply a limiter of density flux: the out-flux density
can not exceed the present electron density of each point. The
physics means of this limiter are clear.

\section{numerical results and the analysis}

W.Cheng's experiment results{\cite{Cheng}} were the best experiment
validate of BRA scheme. In this experiment, the author amplified the
seed pulse in a gas-jet plasma. In the plasma with 2mm effective
length, the intensities of seed pulses were amplified several
thousands times and the length of seed pulses was compressed from
550fs (origin) to about 150fs. The recorded intensity reached
$1.7\times 10^{15} W/cm^2$ which was higher than pump intensity by
more than 1 order in magnitude. The author recorded the broadened
(~800 fs ) and compressed ($<200fs$) output pulses both in the
experiment. Which informed that the amplifying process had been in
non-linear region.

We investigated the process numerically. As the experiments setting
up, we set the seed pulse length to 550 fs and set the pumping pulse
length to 10 ps. The plasma's length was about 2 $mm$ and the
plasma's density was $1.1*10^{19}cm^{-3}$ . The wavelength of
pumping pulse was 0.8 $\mu m$ and the wavelength of seed pulse was
set to matching with the plasma and pumping pulse.

At the peak top of the pumping pulse,we let the two pulses collide
each other. For the uncertainty of the collide points and the plasma
temperatures, the simulating results varied a little. When the
interaction finished, the seed pulse evolved to narrow peek which
FWHM about $150fs-170fs$ and peak intensity about $1.5*10^{15}W/cm^2
-1.9*10^{15}W/cm^2$. The typical simulate results were shown at
figure 1. We can find that the $\pi$-pulse shape easily. Which means
the non-linear amplifying region was reached. Considered the
uncertainly of the plasma temperature and length, we can find good
agreement between the simulations and the experiments.

\begin{figure}
\includegraphics{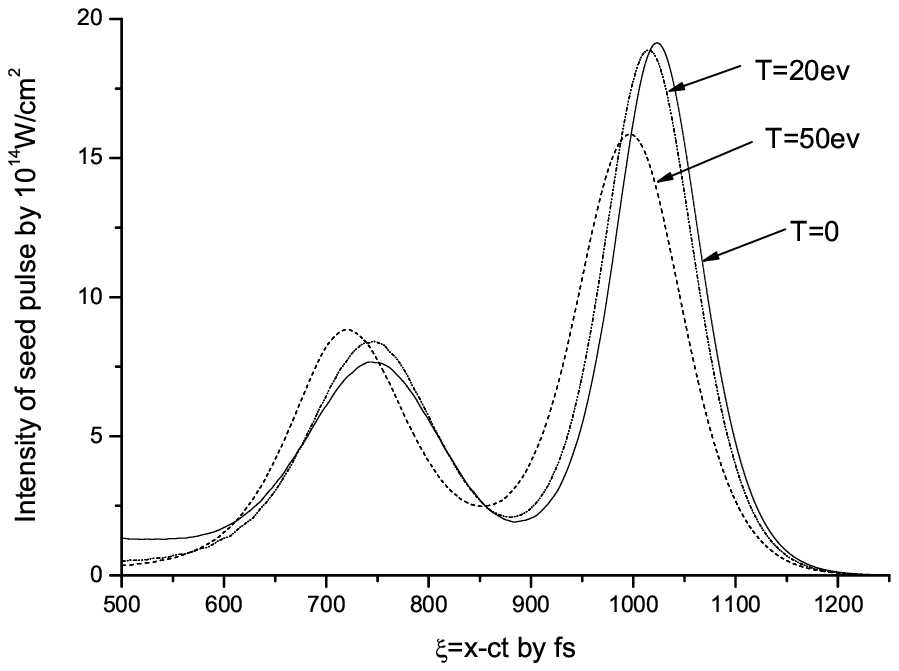}
\begin{center}
figure 1:The output intensity versus $\xi$ after amplified.
\end{center}
\end{figure}

From the simulation results, we can find the evolving process of
seed pulse. Let the seed pulse collide with different length pumping
pulse,we can image the evolving  of the seed pulse: it was broadened
to about 750 fs and compressed later. At the end, it evolved to a
pulse shorter than 160 fs.

The typical results are shown at figure 2.

\begin{figure}
\includegraphics{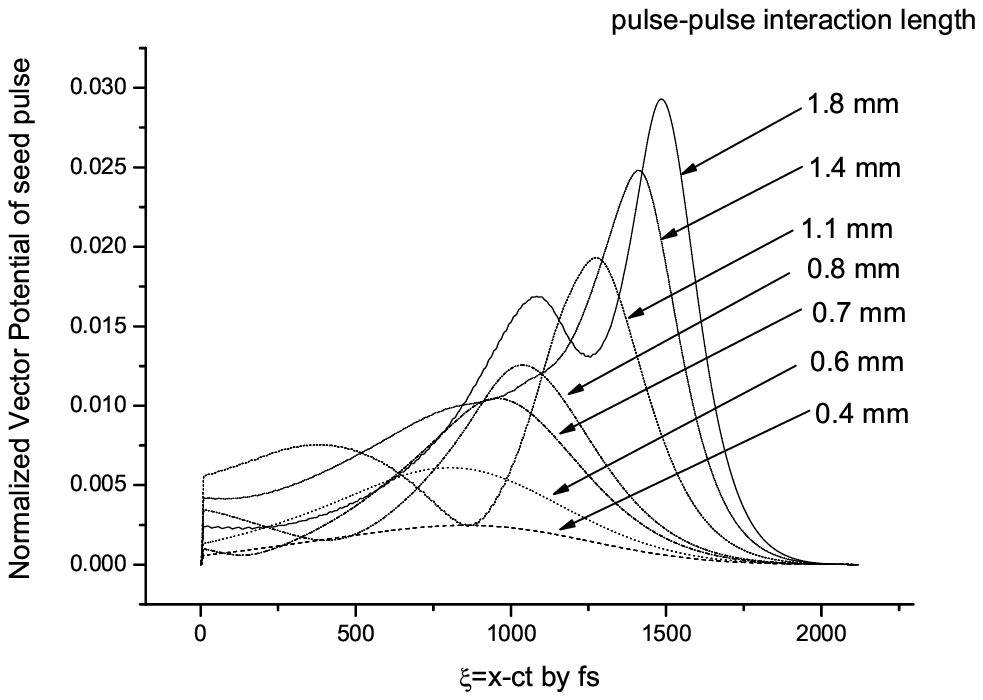}
\begin{center}
figure 2: The amplified pulse shape of the seed by different length
pumping pulse.
\end{center}
\end{figure}

The simulating process shows the non-linear character of plasma
wave. When the amplitude of plasma wave is quite large,the wave's
shape are different from sine wave, the typical wave form are shown
in figure 3. As we see, the difference are quite evidence in our
results. In fact, the effective $\delta n/n$ can exceed 1 when the
amplifier reaches the exhaust region. This phenomena made an adding
stability of amplifier from wave-breaking.
\begin{figure}
\includegraphics{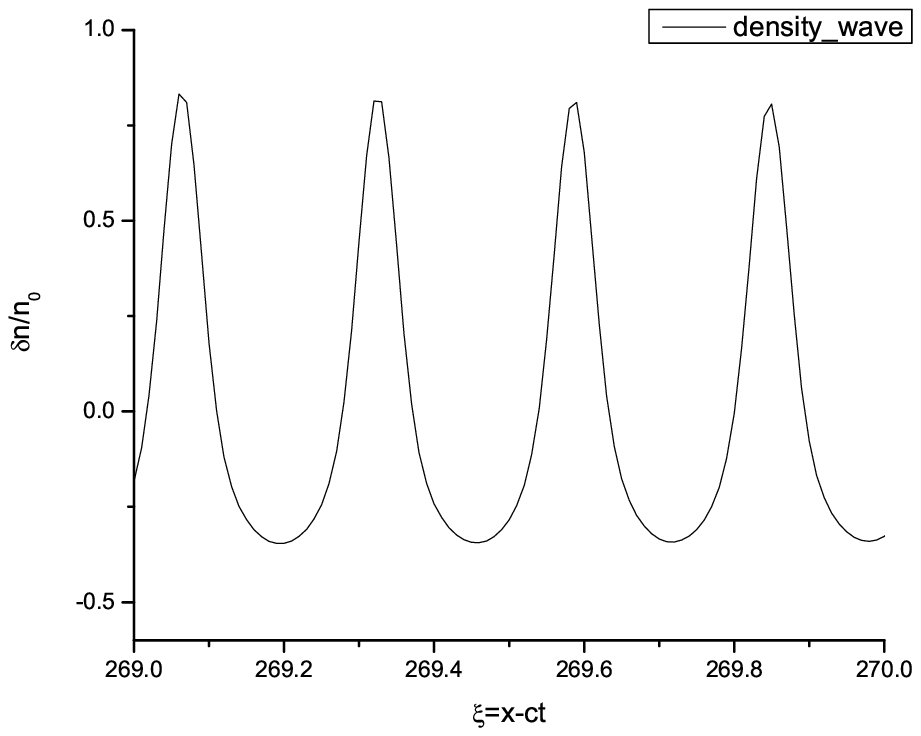}
\begin{center}
figure 3: The typical wave-shape of density oscillating when the
amplitudes become large.
\end{center}
\end{figure}

However, our results show that when the amplifying length reach to
2mm, the plasma wave amplitude was close to cold plasma
wave-breaking amplitude. Which means, if we want to make an
practical amplifier ,we must increase the plasma density to get
higher stabilities. Although It was very difficult to get dense and
uniform plasma in experiments, we simulated the BRA process in a
quite dense plasma ($n=0.04 n_{cr}$). The results were coarse in
precision but show that the amplifying processes were much more
stable and easier to reach the nonlinear-regime in a more dense
plasma. The higher increase rates of raman effects and higher
stability compensated the lower conversion effects in the more dense
plasmas. The results of our simulation are shown in figure 4. In
this density, we need shorter than 1mm plasmas to generate obviously
amplified intensities. It means that the increasing of uniform
plasma density should be the most important improvement at future
experiments.
\begin{figure}
\includegraphics{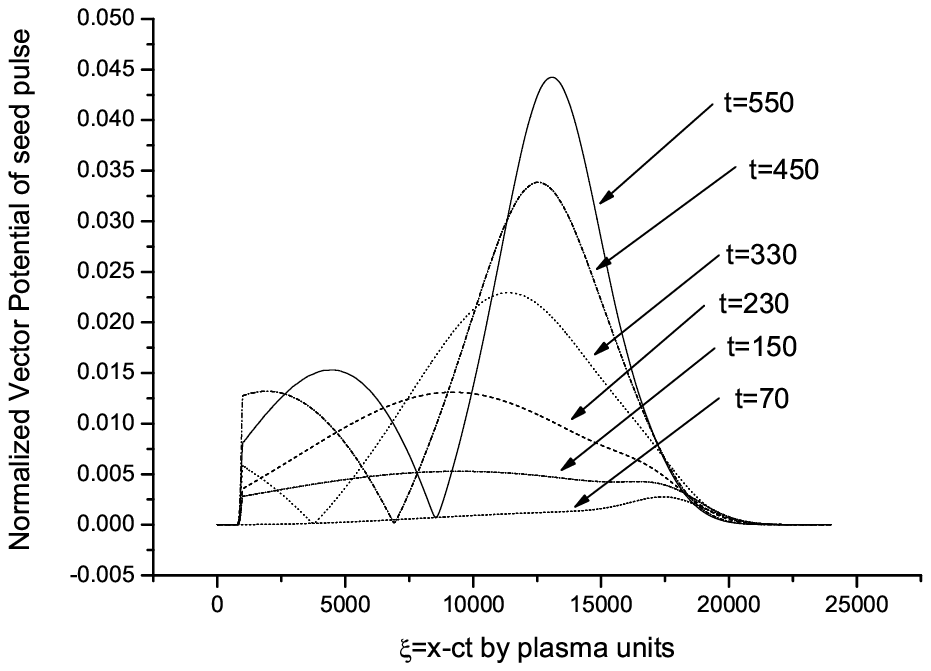}
\begin{center}
figure 4: The pulse shape of the seed after being amplified by
different length pumping pulse;the unit is the natural of plasma.
\end{center}
\end{figure}

Our method can be easily improved to include the effects as pumping
heat,plasma un-uniformity, inverse-Bremsstrahlung, or even linear
landau damping, tunnel-ionized (by phenomena) effects et al. The
investigation of these effects in BRA are proceeding now.

\section{conclusion}
We introduced a hydrodynamic simulation method which was suitable
for the interaction of lasers and corona plasmas. The method is
convenient for the simulation of ultra-fast laser's transmitting. We
simulated the Backward Raman Amplifiers experiments numerically, and
the results show good agreement with  experiments. The following
study will appear in another paper.

\section{acknowledge}
Prof. Chen Bao-Zhen has given some constructive suggestions about
this work..

\end{document}